\documentclass[aps,pre,twocolumn,superscriptaddress]{revtex4}

\usepackage{color}
\usepackage{amsmath}    
\usepackage{amssymb} 
\usepackage{amsfonts}   
\usepackage{graphicx}
\usepackage{pgfplots}
\usepackage{bm}

\usepackage{multirow} 
\usepackage{bbm}
\graphicspath{{/.appendixgr/}}

\setcitestyle{authoryear,open={(},close={)}}

\newcommand{\beq}{\begin{eqnarray}}
\newcommand{\eeq}{\end{eqnarray}}
\newcommand{\<}{\langle}
\renewcommand{\>}{\rangle} 

\newcommand{\s}{\sigma}

\newcommand{\bs}{{\bm \sigma}}

 \newcommand{\CG}[1]{#1}

\begin{document}
\title{A tractable method for describing complex couplings \\ between
  neurons and population rate}

\author{Christophe Gardella}
\affiliation{Laboratoire de physique statistique, CNRS, UPMC and \'Ecole normale sup\'erieure, 24, rue Lhomond, 75005 Paris, France}
\affiliation{Institut de la Vision, INSERM and UMPC, 17 rue Moreau, 75012 Paris, France}
\author{Olivier Marre}
\affiliation{Institut de la Vision, INSERM and UMPC, 17 rue Moreau, 75012 Paris, France}
\author{Thierry Mora}
\affiliation{Laboratoire de physique statistique, CNRS, UPMC and \'Ecole normale sup\'erieure, 24, rue Lhomond, 75005 Paris, France}

\begin{abstract}
 % abstract_6
Neurons within a population are strongly correlated, but how to simply capture these correlations is still a matter of debate. Recent studies have shown that the activity of each cell is influenced by the population rate, defined as the summed activity of all neurons in the population. 
However, 
an explicit,
tractable model for these interactions is still lacking.
Here we build a probabilistic model of population activity that reproduces the firing rate of each cell, the distribution of the population rate, and the \CG{linear} coupling between them.
\CG{
This model is tractable, meaning that its parameters can be learned in
a few seconds on a standard computer even for large population recordings.
}
We inferred our model for a population of 160 neurons in the salamander retina. In this population, single-cell firing rates depended in unexpected ways on the population rate. In particular, some cells had a preferred population rate at which they were most likely to fire. 
These complex dependencies could not be explained by a linear coupling between the cell and the population rate. We designed a more general, still tractable model that could fully account for these non-linear dependencies.		
We thus provide a simple and computationally tractable way 
to learn models that reproduce the dependence of each neuron on 
the population rate.

\end{abstract}

\maketitle

\noindent\fbox{
\begin{minipage}{.95\linewidth}
\vskip .3cm
\section*{Significance statement}
 % significance_6
The description of the correlated activity of large populations of neurons
is essential to understand how the brain performs computations and
encodes sensory information.
These correlations can manifest themselves in the coupling of single
cells to the total firing rate of the
surrounding population, as was recently demonstrated in the
visual cortex, but how to build this dependence into an explicit
model of the population activity is an open question. 
Here we introduce a general and tractable model based on the principle
of maximum entropy to describe this population coupling. By applying our approach to
multi-electrode recordings of retinal ganglion cells, we find complex forms
of coupling, with the unexpected tuning of many neurons to a preferred
population rate.

\vskip .3cm
\end{minipage}
}

\bigskip

 % introduction_6
An important feature of neural population codes is the correlated firing of neurons. Manifestations of collective activity are observed in the correlated firing of individual pairs of neurons \citep{Arnett1978}, and through the coupling of single neurons to the activity in its surrounding population \citep{Arieli1996,Tsodyks1999}.
These correlations, whether they are evoked by common inputs or result from interactions between neurons, imply that the neural code must be studied through the collective patterns of activity rather than by individual neuron.     
  
As the number of possible firing patterns in a population grows exponentially with its size, they cannot be sampled exhaustively for large populations. 
Several modeling approaches have been suggested to describe the collective activity patterns of neural population \citep{Grun1995,Schneidman2003,Schneidman2006,Pillow2008,Cocco2009a,Tkacik2014a}. In these approaches, a small number of statistics (e.g. mean firing rate, pairwise correlations) is measured to constrain the parameters of the model. Models are then evaluated on their ability to predict statistics of the population activity that were not fitted to the data. These models are computationally hard to infer, and one must usually have recourse to approximate methods to fit them.

Recently, \cite{Okun2015} investigated how the activity of the whole population influenced the behavior of single neurons in the primary visual cortex of awake mice and monkeys. In particular, they studied the role of the correlation between neurons and the summed activity of the population, called the population rate. To assess whether these couplings between neurons and population activity were sufficient to describe the correlative structure of the code, synthetic spike trains preserving these couplings were generated and compared to data. However, the numerical method used to generate synthetic spike trains is computationally heavy, and is unable to predict the probability of particular patterns of spikes, as most of them are unlikely to ever occur. 

Here we introduce a new method, based on the principle of maximum entropy, to exactly account for the coupling between individual neurons and the population rate.
\CG{This model is tractable, meaning that predictions for the statistics of the activity can be derived analytically. The gradient and Hessian of the model's likelihood can thus also be computed efficiently, allowing for fast inference using Newton's method.}
 Compared to previous methods \citep{Okun2015}, our method can fit hours of large-scale recordings of large populations in a few seconds on a standard laptop computer. 
 We tested it on recordings of the salamander retina (160 neurons). We uncovered new ways for individual neurons to be coupled to the population, where a single neuron is tuned to a particular value of the population rate, rather than being monotonically coupled to the population.

 % methods_6
\section{Materials and Methods}
\subsection{Recordings from retinal ganglion cells}
We analyzed previously published {\em ex vivo} recordings from  retinal ganglion cells of the tiger salamander (Ambystoma tigrinum) \citep{Tkacik2014a}. In brief, animals were euthanized according to institutional animal care standards. The retina was extracted from the animal, maintained in an oxygenated Ringer solution, and recorded on the ganglion cell side with a 252 electrode array. Spike sorting was done with a custom software \citep{Marre2012}, and $N=160$ neurons were selected for the stability of their spike waveforms and firing rates, and the lack of refractory period violation.

\subsection{Maximum entropy models}
We are interested in modeling the probability distribution of population responses in the retina. The responses are first binned into 20-ms time intervals. The response of neuron $i$ in a given interval is represented by a binary variable $\s_i$, which takes value 1 if the neuron spikes in this interval, and 0 if it is silent. The population response in this interval is represented by the vector $\bs=(\s_1,\ldots,\s_N)$ of all neuron responses (Figure ~\ref{fig:1}A). We define the population rate $K$ as the number of neurons spiking in the interval: $K(\bs) = \sum_{i=1}^N \s_i$.

We build three models for the probability of responses, $P(\bs)$. These models reproduce some chosen statistics, meaning that these statistics have same value in the model and in empirical data. The first model reproduces the firing rate of each neuron and the distribution of the population rate. The second model also reproduces the correlation between each neuron and the population rate. The third model reproduces the whole joint probability of single neurons with the population rate. It is a hierarchy of models, because the statistics of each model are also captured by the next one. \\

{\em Minimal model.}
We build a first model that reproduces the firing rate of each neuron, $P(\s_i\!=\!1)=\<\s_i\>$, and the distribution of the population rate, $P(K)$. We also want the model to have no additional constraints, and thus be as random as possible. In statistical physics and information theory, the randomness of a distribution $P$ is measured by its entropy $S(P)$:
\beq
S(P) = - \sum_\bs P(\bs) \; \ln P(\bs),
\eeq
where the sum runs over all possible states. The maximum entropy model is the distribution that maximizes this entropy while reproducing the constrained statistics. Using the technique of Lagrange multipliers (see Mathematical derivations), one shows that the model must take the form:
\begin{equation}
P(\bs) = \dfrac{1}{Z} \exp\left( \sum_{i=1}^N (\alpha_i + \beta_{K(\bs)}) \; \s_i  \right),
\label{eqn:intermediate_model1}
\end{equation}		
where the parameters $\alpha_i$, $i\!=\!1,\ldots,N$ and $\beta_K$, $K=0,\ldots,N$ must be fitted so that the distribution of Eq.~\ref{eqn:intermediate_model1} matches the statistics $\<\s_i\>$ and $P(K)$ of the data. $Z$ is a normalization factor. \CG{Note that $\beta_{K(\bs)}$ depends on the state $\bs$ through $K(\bs)$.} We refer to this distribution as the minimal model, as no explicit dependency between the activity of individual neurons and the population rate is constrained. \\

{\em Linear-coupling model.}
The second  model reproduces $\<\s_i\>$ and $P(K)$ as before, as well as the linear correlation $\<K\cdot \s_i\>$ between each neuron response $\s_i$ and the population rate $K$, for $i=1,\ldots,N$. It takes the form (see Mathematical derivations):
\begin{align}
P(\bs) &= \dfrac{1}{Z} \exp\left( \sum_{i=1}^N ( \alpha_i + \beta_{K(\bs)} + \gamma_i K) \; \s_i  \right).
\label{eqn:linear_model}
\end{align}
Analogously to the minimal model, the parameters $\alpha_i$, $\beta_K$ and $\gamma_i$ are inferred so that the model agrees with the mean statistics $\< \s_i\>$, $P(K)$ and $\<K\cdot \s_i\>$ of the data. Importantly, despite their common notation, the values of the fitted parameters $\alpha_i$ and $\beta_K$ are different from the ones fitted in the minimal model (see Mathematical derivations). \\

{\em Complete coupling model.}
The third model reproduces the joint probability distributions of the response of each neuron and the population rate, $P(\s_i, K)$. It takes the form (see Mathematical derivations):
\begin{align}\label{eq:ModelDef_methods}
 P(\bs) &= \dfrac{1}{Z} \exp \left(\sum_{i=1}^N h_{iK(\bs)} \: \s_i \right).
\end{align}
The parameters $(h_{iK})_{i=1,\ldots,N;K=0,\ldots,N}$ are inferred so that the model agrees with the data on $P(\s_i, K)$ for each $(i,K)$ pair. \CG{Note that $h_{i K(\bs)}$ depends on the state $\bs$.} We refer to this model as the complete coupling model since it reproduces exactly the joint probability between each neuron and the population rate.

\subsection{Model solution}
The minimal and linear-coupling models can be written in the same form as the complete coupling model (Eq.~\ref{eq:ModelDef_methods}),
but with constraints on the form of $h_{iK}$. In the minimal model, the matrix $h_{iK}$ is constrained to have the form $\alpha_i + \beta_K $. In the linear-coupling model, it is constrained to have the form $\alpha_i + \beta_K + \gamma_i K$.  In the complete coupling model, the matrix $h_{iK}$ has no imposed structure, and all its elements must be learned from the data.

Since all the considered models can be viewed as sub-cases of the complete coupling model (Eq.~\ref{eq:ModelDef_methods}), we only describe the mathematical solution to this general case. First we describe how to solve the {\em direct} problem, {\em i.e.} how to compute statistics of interest, such as $P(\sigma_i,K)$, from the parameters $h_{iK}$. In the next section, we explain how to solve the {\em inverse} problem -- the reverse task of inferring the model parameters from the statistics -- which relies on the solution to the direct problem.

A model is considered tractable if there exists an analytical expression for the normalization factor,
\beq\label{Z}
Z = \sum_{\bs} \exp\left( \sum_{i=1}^N h_{iK(\s)} \s_i \right),
\eeq
allowing for its rapid computation, e.g. in polynomial time in $N$.
All statistics of the model, such as $P(\sigma_i,K)$ or covariances $\<\sigma_i\sigma_j\>-\<\sigma_i\>\<\sigma_j\>$ between pairs of neurons, can then be calculated efficiently through derivatives of $Z$ (see Mathematical derivations). 
In general maximum entropy models are not tractable, because sums of the kind in Eq.~\ref{Z} involve a sum over an exponential number of terms ($2^N$).
Fortunately, in our case, the technique of probability-generating functions provides an expression for $Z$ which is amenable to fast computation using polynomial algebra (see Mathematical derivations):
\beq
Z=\sum_{K=0}^N \mathrm{Coeff}\left[\prod_{i=1}^N (1+Xe^{h_{iK(\s)}}),X^K\right],
\eeq
where $\mathrm{Coeff}[Q,X^n]$ denotes the coefficient of polynomial $Q$ of order $X^n$.

\subsection{Model inference}
We now describe how to fit the models to experimental data. The inference of the model parameters  is equivalent to a problem of likelihood maximization \citep{Ackley1988}. The model reproduces the empirical statistics exactly when the parameters maximize the likelihood of experimental data measured  by the model, $L=\prod_{\alpha=1}^n P(\bs^{(\alpha)})$, where $(\bs^{(1)},\ldots,\bs^{(n)})$ are the $n$ activity patterns recorded in the experiment, assumed to be independently drawn.

In practice, we maximized the normalized log-likelihood $\mathcal{L} = (1/n) \log L$ instead of $L$, which is equivalent theoretically but more convenient for computation. We used Newton's method  to perform the maximization. This method requires to compute the first and second derivatives of the normalized log-likelihood. These derivatives can be expressed as functions of mean statistics of the model and can be calculated using the solution to the direct problem sketched in the previous section, and detailed in the Mathematical derivations. Because the model is tractable, these mean statistics can be computed quickly and the model can be inferred rapidly.

For the minimal model, the optimization was performed over the parameters $(\alpha_i)_{i=1,\ldots,N}$ and  $(\beta_K)_{K=0,\ldots,N}$. For the linear-coupling model, the optimization was done over these two sets of parameters, as well as $(\gamma_i)_{i=1,\ldots,N}$.
For the complete coupling model, the optimization was performed over all elements of the matrix $(h_{iK}) _{i=1,\ldots,N;K=0,\ldots,N}$. We stopped the algorithm when the fitting error was smaller than $10^{-6}$ (see Mathematical derivations).

\subsection{Regularization}
Prior to learning the model, we regularized the empirical population rate distribution $P(K)$ and conditional firing rates $P(\s_i|K)$ to mitigate the effects of low-sampling noise. This regularization allowed us to remove zeros from the mean statistics, avoiding issues with the fitting procedure.
We performed this regularization using pseudocounts (see Mathematical derivations).

\subsection{Tuning curves in the population rate}
We define the tuning curve of neuron $i$ in the population rate as the conditional probability of neuron $i$ to spike given the summed activity of all neurons but $i$, $K_{\backslash i} \!\!=\!\! \sum_{j\neq i} \s_j$. It is equal to:
\beq
 P(\s_i\!\!=\!\!1|K_{\backslash i}) =\frac{P(\s_i\!\!=\!\!1,K_{\backslash i})}{P(\s_i\!\!=\!\!0,K_{\backslash i})+P(\s_i\!\!=\!\!1,K_{\backslash i})}
\eeq
where we can use $P(\s_i\!\!=\!\!1,K_{\backslash i}) = P(\s_i\!\!=\!\!1,K\!\!=\!\!K_{\backslash i}\!\!+\!\!1)$ and $P(\s_i\!\!=\!\!0,K_{\backslash i}) = P(\s_i\!\!=\!\!1,K\!\!=\!\!K_{\backslash i})$. Each of these quantities can be computed using the solution to the direct problem (see Mathematical derivations).

We then tested for each neuron if its tuning curve had significant local maxima. We first identified the set of $K_{\backslash i}$ for which $P(\s_i=1|K_{\backslash i})$ was significantly larger than points below and above $K_{\backslash i}$. To assess significance, we measured the standard deviation of the difference across 100 training sets consisting in random halves of the dataset. The difference was said to be significant when it was 5 standard deviations above 0.

For the cells for which the presence of a maximum was determined, we then evaluated the location of the maximum, $K_{\backslash i}^*$, by taking the median of the maxima determined for each training set. We inferred the presence and position of minima in a similar way.

To estimate the quality of the model prediction for the tuning curve, we quantified how the model differed from the data. We trained the model on 100 random training sets and computed  $D_{KL}({\rm test}\|{\rm model})$, the difference between $P(\s_i,K)$ in the testing data and predicted by the model, measured by the Kullback Leibler (KL) divergence. The KL divergence between two distributions $P$ and $Q$ of a random variable $x$ is: $ D_{KL}(P \| Q) = \sum_x P(x) \text{log}[{P(x)}/{Q(x)}]$. We regularized $P(\s_i,K)$ in the testing set before computing the KL divergence.
To measure sampling noise we computed the difference between the testing and the training sets,  $D_{KL}($test$\|$train$)$, where $P(\s_i, K)$ was regularized in both sets. The normalized KL divergence, $z,$ is defined as the difference between $D_{KL}($test$\|$model$)$ and $D_{KL}($test$\|$train$)$, divided by the standard deviation:
\begin{equation}
 z = \dfrac{\text{mean}\left(D_{KL}(\text{test}\|\text{model}) - D_{KL}(\text{test}\|\text{train}) \right)}{\text{std}\left(D_{KL}(\text{test}\|\text{model}) - D_{KL}(\text{test}\|\text{train}) \right)}.
\end{equation}
In other words, it measures by how many standard deviations the data differs from the model.

\subsection{Quality of the model}
{\em Pairwise correlations.}
In order to measure the quality of the predictions of 
correlations between pairs of neurons $\s_i$ and $\s_j$, we used cross-validation. We randomly divided the dataset into 100 training and testing sets half the size of data, and learned the model on the training sets. The correlations of each testing set, $c_{{\rm test},\, ij}$ were then predicted with the model $c_{{\rm model},\,ij }$. The quality of the model prediction was measured \CG{by a goodness-of-fit index quantifying the amount of correlations predicted by the model. We define it as:}
\begin{equation} \label{eq:explained_cc}
 C= \dfrac{\sum_{i<j}c_{{\rm test},ij }^2 - \sum_{i<j}\left(c_{{\rm test},ij } - c_{{\rm model},\,ij }\right)^2}{\sum_{i<j}c_{{\rm test},ij }^2 - \sum_{i<j}\left(c_{{\rm test},ij } - c_{{\rm train},ij }\right)^2},
\end{equation}
where $c_{{\rm train},\, ij}$ is the correlation in the corresponding training set. The numerator of Eq.~\eqref{eq:explained_cc} is the part of the correlations in the testing set that is predicted by the model, and the lower one is a normalization correcting for sampling noise. We have $C=1$ when the model perfectly accounts for the correlations of the training set. When the model completely ignores correlations, as in a model of independent neurons, $c_{ij}=0$, then $C=0$. \\

{\em Likelihood.}
Using the models learned on the same 100 training sets, we computed the likelihood of responses in the testing sets for the minimal, linear-coupling and complete coupling models. In this paper the log-likelihood is expressed in bits, using binary logarithms.
We then computed the improvement in mean log-likelihood compared to the minimal model as the ratio $\< \, \< \log P(\bs)\>_\bs \, / \, \< \log P(\bs) \>_\bs \, \>_\text{test}$, where $\< \cdot \>_{\text{test}}$ is the mean over training sets and $\< \cdot \>_{\bs}$ is the mean over responses in each testing set.\\

{\em Multi-information.}
The multi-information \citep{Cover,Schneidman2003} quantifies the amount of correlative structure captured by a model. It is defined as the difference between the entropy of a model of independent neurons reproducing firing rates, and empirical data: $I = S_{\rm indep} - S_{\rm data}$. Here $S_{\rm data}=-\sum_\bs P_{\rm data}(\bs)\log P_{\rm data}(\bs)$ is the entropy of the spike patterns measured by their frequencies $P_{\rm data}(\bs)$ in the data, and $S_{\rm indep}$ is the entropy if all neurons were independent.

The entropy of a maximum entropy model is by construction higher than that of the real data, $S_{\rm model}>S_{\rm data}$, because the model has maximum entropy given the statistics it reproduces. Its entropy is also smaller than $S_{\rm indep}$ provided that constraints include the spike rates, because the model has more structure, reproduces more statistics, than if neurons were independent, Thus the fraction of correlations that is accounted by the maximum entropy model, $0<I_{\rm model}/I<1$, where $I_{\rm model}=S_{\rm indep}-S_{\rm model}$, can be viewed as a measure of how well the model captures the correlative structure of responses. The true multi-information $I$ can only be calculated for small groups of neurons ($N\leq 20$), because $P_{\rm data}$ requires to evaluate $2^N$ pattern frequencies, which is prohibitive for large networks.

 % results_6
\section{Results}
\subsection{Tractable maximum entropy model for coupling neuron firing to population activity}

\begin{figure}
\centering
\includegraphics[width=\linewidth]{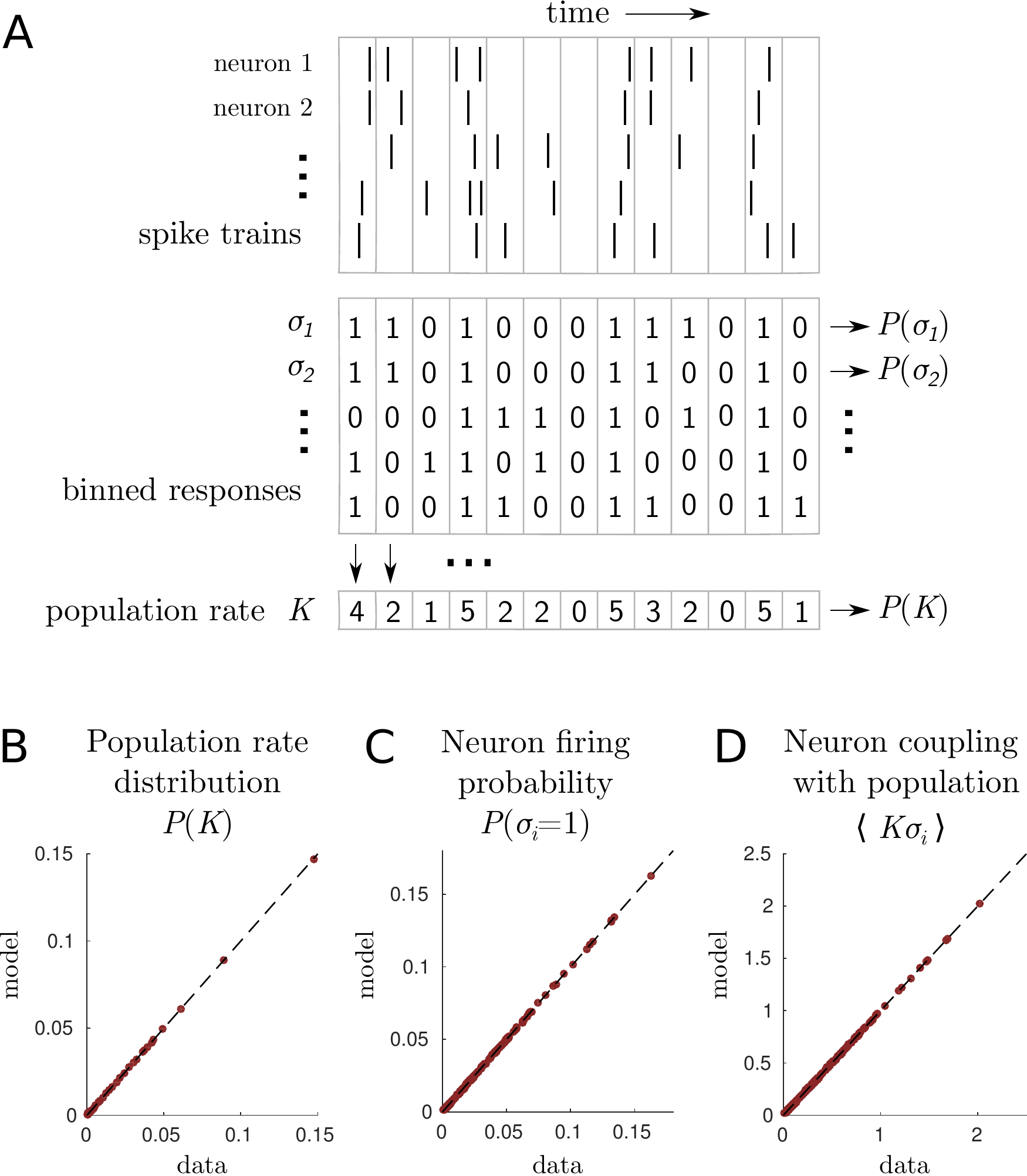}
\caption{\textbf{A maximum entropy model for population coupling.} \textbf{A}, Spikes trains are recorded with a multieletrode array and binned in 20ms time windows. We study the dependence between each neuron's binned response, $\s_i$, and the population rate $K$, defined as the summed activity of all neurons. \textbf{B-D}, The linear-coupling model fits three observables with high accuracy: the population rate distribution (B), the cells firing rates (C), and population couplings (D). For each observable the model fit is plotted against empirical values.}
\label{fig:1}
\end{figure}

The principle of maximum entropy \citep{Jaynes1957a,Jaynes1957} provides a powerful tool to explicitly construct probability distributions that reproduce key statistics of the data, but are otherwise as random as possible.
We introduce a novel family of maximum entropy models of spike patterns that preserve
 the firing rate of each neuron, the distribution of the population rate, and the correlation between each neuron and the population rate, with no additional assumptions (Figure \ref{fig:1}A). 
 Under these constraints, the maximum entropy distribution over spike patterns in a fixed 20-ms time window is given by (see Materials and Methods):
\begin{equation}
\label{ModelDef_linear}
 P(\sigma_1,\ldots,\sigma_N) = \frac{1}{Z} \exp\left( \sum_{i=1}^N ( \alpha_i + \beta_K + \gamma_i K)\,  \s_i \right)
\end{equation}
where $\s_i$ equals 1 when neuron $i$ spikes within the time window, and 0 otherwise, $K = \sum_i \s_i$ is the population rate, and $Z$ is a normalization constant. The parameters $(\alpha_i)_{i=1,\ldots,N}$, $(\gamma_i)_{i=1,\ldots,N}$ and $(\beta_K)_{K=0,\ldots,N}$ must be fitted to empirical data. We refer to this model as the linear-coupling model, because of the linear term $\gamma_i K \s_i$ in the exponential.

Unlike maximum entropy models in general, this model is tractable, meaning that its prediction for the statistics of spike patterns has an analytical expression that can be computed efficiently using polynomial algebra. This allows us to infer the model parameters rapidly for large populations on a standard computer, using Newton's method (see Materials and Methods). We learned this model in the case of a population of $N=160$ salamander retinal ganglion cells, stimulated by a natural movie.
It took our algorithm 14 seconds to fit the $3N-2$ model parameters (see Mathematical derivations) so that the maximum discrepancy between the model and the data was smaller than $10^{-6}$ (Figure~\ref{fig:1} B-D).

The linear-coupling model provides a rigorous mathematical formulation to the hypotheses underlying the modeling approach of \cite{Okun2015} applied to cortical populations. In that work, synthetic spike trains were generated by shuffling spikes from the original data so as to match the three constraints listed above on the single-neuron spike rates, the distribution of population rates, and their linear correlation. Shuffling data, {\em i.e.} increasing randomness and hence entropy, while constraining mean statistics, has previously been shown to be equivalent to the principle of maximum entropy in the context of pairwise correlations \citep{Bialek2008a}. Our formulation provides a fast way to learn the model and to make predictions from it, as we shall see below. In addition, it allows us to calculate the probability of individual spike patterns, Eq.~\ref{ModelDef_linear}, which a generative procedure such as the one in \cite{Okun2015} cannot. 

 % figure1_3

\subsection{Tuning curves of single neurons to the population activity}
\begin{figure}
\centering
\includegraphics[width=\linewidth]{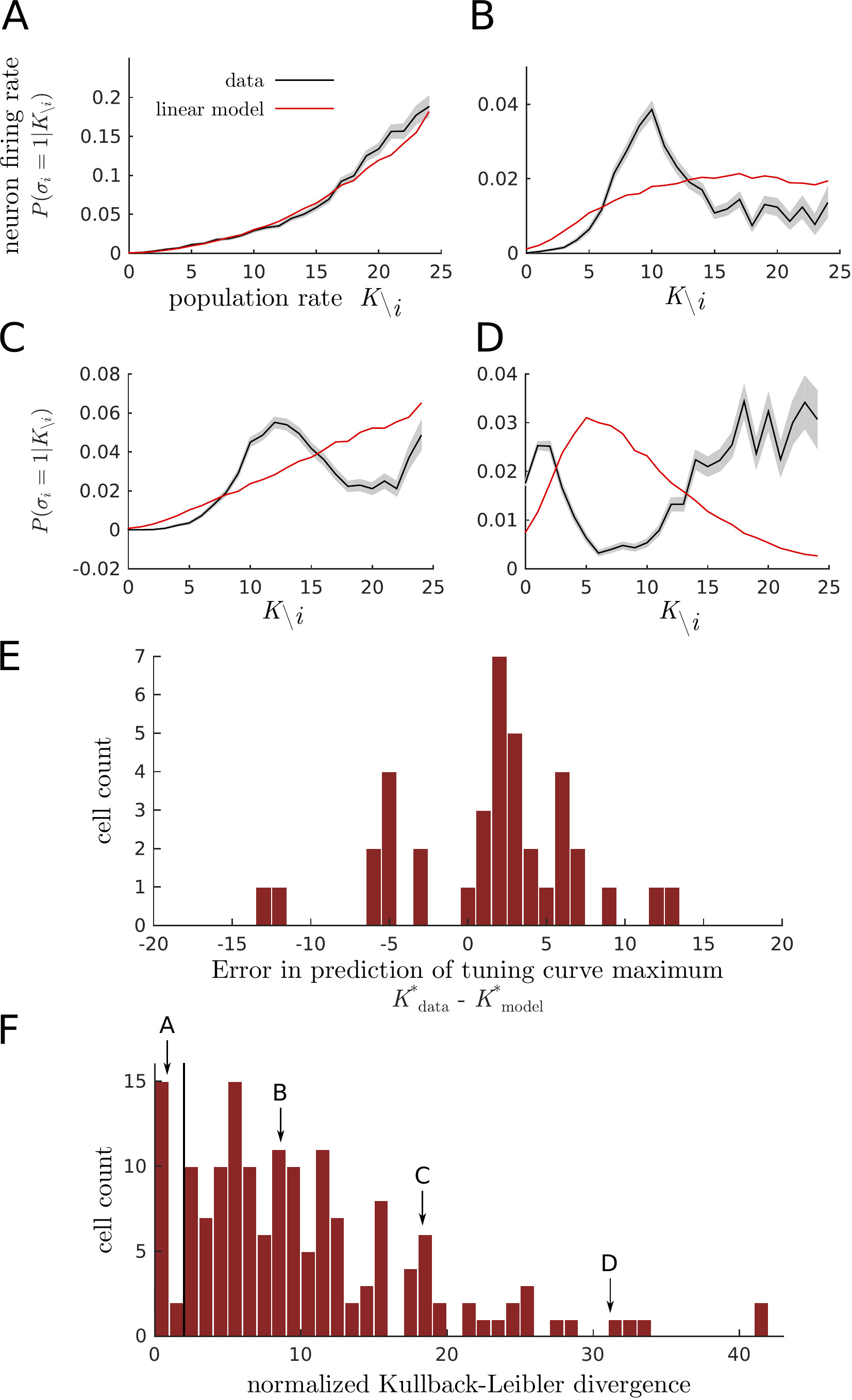}
\caption{\textbf{Tuning curves of single neurons as a function of the population rate.} \textbf{A-D}, Spiking probability of neuron $i$ conditioned on the summed activity of all other neurons, $P(\s_i =1 | K_{\backslash i})$, as observed in the data (black curves, standard error shaded in grey), and predicted by linear-coupling model (red curves). 
Each subfigure corresponds to a different representative cell. 
\textbf{E}, Histogram of the difference between the preferred population rate -- at which the tuning curve is maximal -- observed in the data, $K_{\rm data}^*$, and predicted by the linear-coupling model, $K_{\rm model}^*$. Data are shown for the 38 cells that had at least one local maximum both in the linear model and in the data. When the empirical tuning curve had 2 local maxima, the closest one to the model prediction was chosen.
\textbf{F}, Histogram of the normalized Kullback-Leibler divergence between the observed joint distributions $P(\s_i, K)$ and its prediction by linear-coupling model.
The arrows indicate the value for the four example cells A-D. The vertical line shows a normalized divergence of 2, meaning that cells sitting on its right deviate from the linear-coupling model by more than 2 standard deviations.}
\label{fig:Pi_condKo_selectcells}
\end{figure}

We wondered whether the linear-coupling model could explain how the response of single neurons depended on the population rate.
We examined the firing probability of neuron~$i$ as a function of the summed activity of the other neurons $K_{\backslash i} = \sum_{j \neq i}\s_j$, denoted by $P(\s_i\!=\!1 | K_{\backslash i})$. 
This quantity can be viewed as the tuning curve of neuron~$i$ in response to the rest of the population.
It can be calculated analytically from the parameters of the model (see Materials and Methods), and compared to empirical values. 
The tuning curves of four representative cells are shown in Figure~\ref{fig:Pi_condKo_selectcells}A-D.

The linear-coupling model predicts a variety of tuning curves  (in red), from sub-linear to super-linear.
Although its prediction was qualitatively close to the empirical value for some cells (Figure ~\ref{fig:Pi_condKo_selectcells}A), the model generally did not account well for the coupling between $\s_i$ and $K_{\backslash i}$. A majority of cells (85 out of 160) displayed a local maximum in their empirical tuning curves, at some prefered value $K_{\backslash i}^*$ of the population activity to which the neuron is tuned. The model did not predict the existence of this maximum in 47 out of these 85 cells (Figure~\ref{fig:Pi_condKo_selectcells}C). Even when it did, the location of the maximum,  $K_{\backslash i}^*$, was poorly predicted, as can be seen by the distribution of the difference between model and data (Figure~\ref{fig:Pi_condKo_selectcells}E). In six cases, the tuning curve had two local maxima, while 
the model only predicted one. Another 27 cells had a minimum in their empirical tuning curve, which was never reproduced by the model (Figure~\ref{fig:Pi_condKo_selectcells}D).
Interestingly, no cells were tuned to fire when the rest of the population is silent; even cells whose spiking activity was anti-correlated to the rest of the population had a non-zero preferred population rate, $K_{\backslash i}^*>0$. 

The model performance can be quantified by computing the Kullback Leibler (KL) divergence between data and model for the joint probability $P(\s_i,K)$ of the neuron and population activity (Figure~\ref{fig:Pi_condKo_selectcells}F). The KL divergence is a measure of the dissimilarity between two distributions $P$ and $Q$ \citep{Cover}, which quantifies the amount of information that is lost if we use $Q$ to approximate $P$.
We calculated a normalized KL divergence (see Materials and Methods) measuring by how many standard deviations the KL divergence 
between linear-coupling model and data
deviated from what would be expected from sampling noise (Figure~\ref{fig:Pi_condKo_selectcells}F). A majority of cells (143 out of 160) deviated by more than 2 standard deviations, meaning that their tuning curve was not well accounted for by the linear-coupling model.
This observation is consistent with the model's failure to account for the qualitative properties of their tuning curves.

Taken together, these results indicate that the full dependency between single cells and the population rate cannot be explained by their linear correlation only.

 % figure2_4

\subsection{A refined maximum entropy model}
To overcome the limitations of the linear-coupling model, and to fully account for the variety of tuning curves found in data, we built a maximum entropy model constrained to match all joint probabilities of the population rate with each single neuron response, $P(\s_i,K)$. This model takes the form (see Materials and Methods):
\begin{equation}
\label{ModelDef_general}
 P(\sigma_1,\ldots,\sigma_N) = \frac{1}{Z} \exp\left( \sum_{i=1}^N h_{iK}  \s_i \right),
\end{equation}
where the parameters $h_{iK}$ for $i=1,\ldots,N$ and $K=0,\ldots,N$ are fitted to empirical data, and $Z$ is a normalization constant.
Note that the linear-coupling model can be viewed as a particular case of this model, with parameters $h_{iK}$ constrained to take the form $h_{iK} = \alpha_i + \beta_K + \gamma_i K$.
By construction, this model exactly reproduces the tuning curves of Figs.~\ref{fig:Pi_condKo_selectcells}A-D.

Although this model has many more parameters than the simpler linear-coupling model, it is still tractable, and we could infer its $N(N-1)+1$ parameters (see Mathematical derivations) in 7 seconds for the whole population of 160 neurons. Hereafter, we refer to this model as the complete coupling model.

\subsection{Pairwise correlations}
\begin{figure}
\centering
\includegraphics[width=1\linewidth]{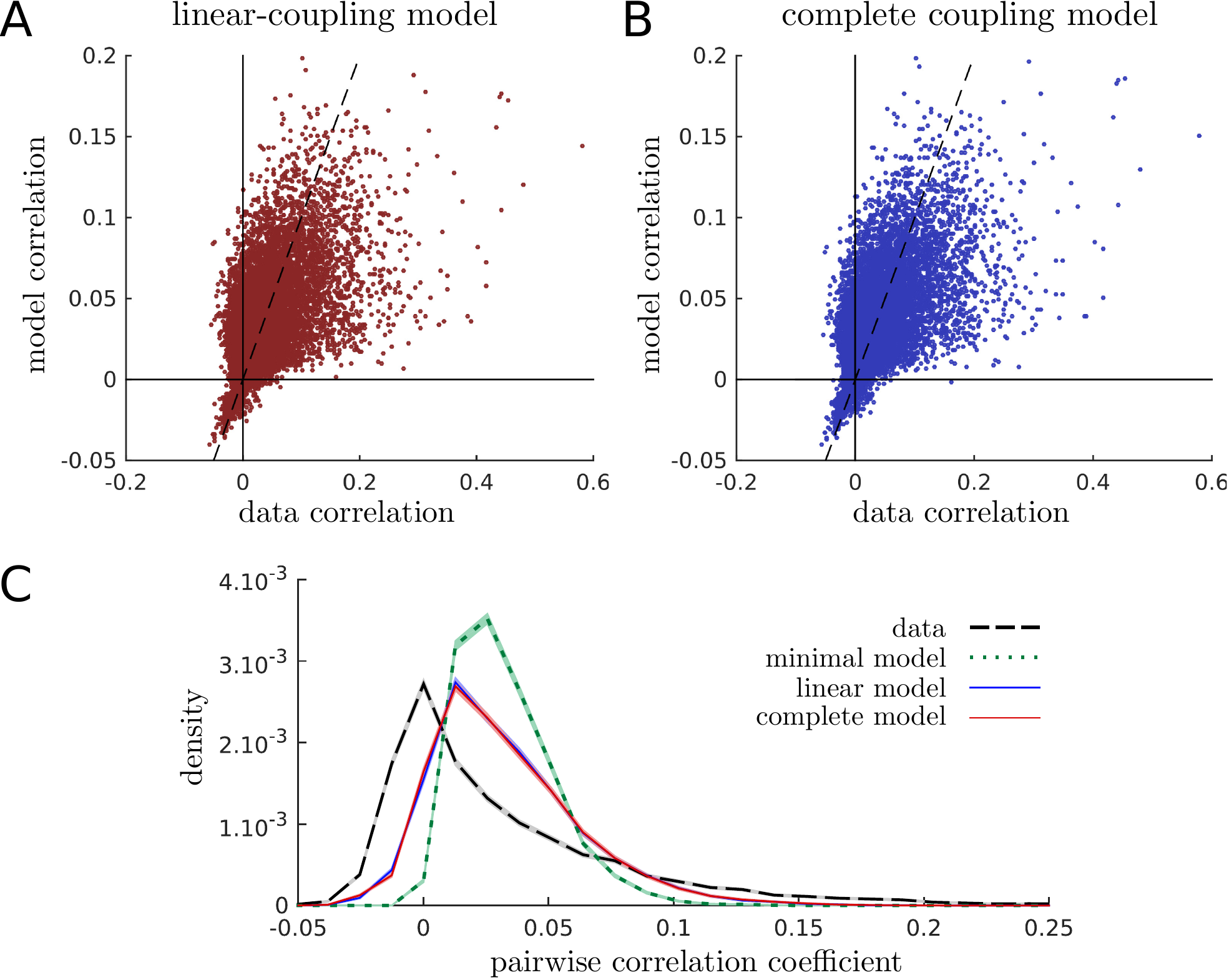}
\caption{\textbf{Maximum entropy models of population coupling partly
    account for pairwise correlations.} \textbf{A-B}, The observed correlation
  coefficient between all pairs of neurons is compared to its prediction according to
  the linear (A) and complete (B) coupling models. \textbf{C},
  Distribution of pairwise correlation coefficients, as observed in the data and predicted by minimal, linear and complete coupling models.}
\label{fig:correlations_hist}
\end{figure}

The models introduced thus far are only constrained to reproduce the firing rate of each neuron, the distribution of population rates, and the coupling of each neuron with the population rate.
We asked whether these simple models
could account for correlations between individual pairs of cells, which were not fitted to the data.
The correlation between two neurons, $\<\s_i\s_j\>-\<\s_i\>\<\s_j\>$, can be calculated analytically from the model parameters (see Materials and Methods) and directly compared to the data (Figure ~\ref{fig:correlations_hist}A and B). 

In order to understand the importance of the population rate coupling for the prediction of pairwise correlations, we built a null model only constrained by the firing rates of each neuron and the distribution of the population rate. This simpler maximum entropy model reads:
\begin{equation}
\label{ModelDef_minimal}
 P(\sigma_1,\ldots,\sigma_N) = \frac{1}{Z} \exp\left( \sum_{i=1}^N (\alpha_i + \beta_K) \,  \s_i \right).
\end{equation}
We call it the minimal model. Interactions between neurons only derive from the fluctuations of the population activity, rather than from an explicit coupling.  This model has $2N-1$ parameters (see Mathematical derivations), which are inferred using the same techniques as before.

To quantify the performance of the different models, we calculated a goodness-of-fit index ranging from 0 when the correlations were not predicted at all to 1 when they were predicted perfectly (Materials and Methods). This index was 0.380 $\pm$ 0.001 for the minimal model, 0.526 $\pm$ 0.002 for the linear-coupling model, and 0.544 $\pm$ 0.002 for the complete coupling model. Thus, a substantial amount of pairwise correlations could be explained from the coupling of neurons to the population. By this measure, the complete model performed slightly (but significantly) better than the linear-coupling model.

Figure~\ref{fig:correlations_hist}C shows the distribution of the pairwise correlations in the data and as predicted by the three models.
The minimal model fails to reproduce the long tail of large correlations, and predicts no negative correlations, while 28\% of empirical correlations are negative. By contrast, the linear and complete coupling models predict 7.6\% and 7.7\% of negative interactions respectively, and a longer tail of large correlation coefficients.
Thus the coupling to the population rate is important to reproduce both large correlations and the strong asymetry of the distribution.
 
 % figure4_3

 % table1_6
\begin{table*}
 \begin{center}

\begin{tabular}{cc|c|c|c|c|} 
\cline{3-6}
& & \multicolumn{1}{ |c| }{Data} & Minimal & Linear & Complete \\ \cline{1-6}
\multicolumn{1}{ |c  }{\multirow{2}{*}{$N= 10$} } &
\multicolumn{1}{ |c| }{$I$} &0.0713  $\pm$    0.0398  &    0.0343  $\pm$    0.0268  &    0.0478  $\pm$    0.0315  &    0.0498  $\pm$    0.0326 \\ \cline{2-6}
\multicolumn{1}{ |c  }{}                        &
\multicolumn{1}{ |c| }{${I}/{I_{\rm data}}$} & 1 & 0.444 $\pm$ 0.135 & 0.649 $\pm$ 0.108  & 0.677 $\pm$ 0.108  \\ \cline{1-6}
\multicolumn{1}{ |c  }{\multirow{2}{*}{$N= 20$} } &
\multicolumn{1}{ |c| }{$I$} &0.3188  $\pm$    0.0967  &    0.1291  $\pm$    0.0558  &    0.1702  $\pm$    0.0592  &    0.1789  $\pm$    0.0606 \\ \cline{2-6}
\multicolumn{1}{ |c  }{}                        &
\multicolumn{1}{ |c| }{${I}/{I_{\rm data}}$} & 1 & 0.393 $\pm$ 0.071 & 0.531 $\pm$ 0.055 & 0.557 $\pm$ 0.055\\ \cline{1-6}
\end{tabular}
\caption{Mean ($\pm$ standard deviation) of the multi-information $I$ estimated either directly from the data, or from maximum entropy models, for random sub-populations of 10 and 20 neurons (100 sub-populations each), as well as the ratio of the multi-information between model and data. Results are in bits.}
\label{table}
\end{center}
\end{table*}

\subsection{Prediction of probabilities of spike patterns}
We quantified the capacity of models to describe population responses by computing the probability of responses predicted by each model. The mean log-likelihood of responses was -33.10 $\pm$  0.06 bits for the minimal model, -30.12  $\pm$  0.06 bits for the linear-coupling model and -29.49 $\pm$  0.06 bits for the complete coupling model. 
The improvement in mean log-likelihood compared to the minimal model was 51.3 $\pm$ 0.5 \% higher for the complete coupling model than for the linear-coupling model, meaning that nonlinear couplings to the population are important to model the probability of responses.

The multi-information $I$ \citep{Cover} quantifies, in bits, the amount of correlations in the response, whether they are pairwise or of higher order (see Materials and Methods). To assess the performance of the models in capturing the collective behavior of the networks, we calculated the ratio of the multi-information explained by the model to that estimated directly from the data, $I_{\rm model}/I_{\rm data}$. This ratio gives a measure of how well the probability of particular spike patterns are predicted by the model:
it is 1 when the model is a perfect description of the data, and 0 when the model assumes independent neurons with no correlation between them. Because it requires to estimate the probability of all possible spike patterns of the populations, the multi-information can only be calculated for small populations of at most 20 cells.

With this measure, the linear coupling model could account for $65\%$ of the multi-information for groups of $10$ neurons, and $53\%$ for groups of 20 neurons. The complete model slightly improved these ratios to $68\%$ and $56\%$, respectively (see Table \ref{table}). Thus, more than half of the correlative structure in the spike patterns could be explained by the coupling to the population rate alone.

\section{Discussion}
 % discussion_6
In this work we have introduced a general computational model for coupling individual neurons to the population rate.
This model formalizes and simplifies the generative procedure proposed by \cite{Okun2015} to study population coupling, and overcomes its computational difficulties. In addition, it allows for non-linear coupling to the population rate. 

We have used our model to investigate population coupling in large recordings of $N=160$ retinal ganglion cells. We found that most cells had a non-linear coupling to the population rate. In particular, a large fraction of cells were tuned to a preferred value of the population rate. Even more strikingly, a few cells had a least preferred population rate, {\em i.e.} they were more likely to spike at lower or higher populations rates. We found no cell that was maximally active when all other neurons were silent, even among cells that were anti-correlated with the population rate.
These results emphasize the need for the non-linear coupling afforded by our model
, as they  uncover new dependencies that do not fit within the proposed division between soloists and choristers \citep{Okun2015}, such as the tuning to a specific population rate. 
It would be interesting to test if these non-linear couplings can also be found at the cortical level. 

Overall, our model reaches a similar predictive performance than what was found in the cortex. The coupling to the population rate accounted for more than half of
correlations between pairs of neurons.
In \cite{Okun2015}, a custom measure of the fraction of explained pairwise correlations (different from the one used in the present work) gave 0.34. Applying the same measure to our case yields a similar value, 0.33. However, this similarity in performance can be due to different underlying mechanisms. In the retina, most correlations are due to common input from previous layers \citep{Trong2008}, while ganglion cells do not make synaptic connections to each other. In contrast, at the cortical level, a larger part of the variability in the activity should be due to internal dynamics generated by recurrent connections \citep{vanVreeswijk1996,Arieli1996,Tsodyks1999}. It would be interesting to test our model on cortical data to see if these differences result in different types of non-linear population coupling. 

Our maximum entropy model of population coupling is complementary to maximum entropy models reproducing correlations between all pairs of neurons. Pairwise models have been shown to accurately describe the collective activity of retinal ganglion cells \citep{Schneidman2006,Shlens2006,Shlens2009,Ganmor2011,Ganmor2011a,Tkacik2014a}, and in cortical networks {\em in vitro} \citep{Tang2008} and {\em in vivo} \citep{Yu2008}, but they are not tractable,
requiring to sum over $2^N$ all possible spiking states in order to implement Boltzmann-machine learning \citep{Ackley1988}. Alternative methods based on mean-field approximations \citep{Cocco2009a,Cocco2011} or Monte-Carlo simulations \citep{Broderick2008} have been proposed. However, Monte-Carlo methods require hours of computations, although recent efforts have tried to lower these computation times for moderately large populations \citep{Ferrari2015}.

By contrast, the models of population couplings introduced here are much easier to solve. 
They are tractable, so their predictions can be computed analytically in time $N^3$, and their parameters can be inferred in a few seconds on a personal computer from large-scale, hour-long recordings of spike trains for a population of $N=160$ neurons. 
These models can then be used to generate synthetic spike trains, to calculate analytically response statistics such as pairwise correlations, or to estimate the probability of particular spike trains. 
Compared to the shuffling method described in \cite{Okun2015}, which is equivalent to the linear-coupling model, our method is simpler and computationally less intensive.

The procedure is general and can be applied to any multi-neuron recording of individual spikes.
The speed of model inference could prove an important advantage when studying very large populations, which can now reach a thousand cells \citep{SchwarzD2014}. 
In the case of the linear-coupling model, the number of
parameters is also smaller, scaling with the population
size $N$ rather than $N^2$ for pairwise correlations model.

Note that the population coupling models introduced here belong to a different class than pairwise models. Each class captures features of neural responses that the other cannot:
models of population  coupling should be sufficient for studying the global properties of collective activity, while pairwise models are still needed to account for the detailed structure of the response statistics. Pairwise models have been reported to capture 90\% of the correlations as measured by the multi-information for populations of size $N=10$ \citep{Schneidman2006}, while our model captures at most $70\%$ (Table I). Yet pairwise models can also miss important aspects of the collective activity such as the probability of large population rates \citep{Tkacik2014a}, which is captured by our population model.

Both classes of models consider same-time spike patterns, with no regard for the dynamics of spike trains and their temporal correlations. Generalizations of pairwise maximum entropy models to temporal statistics are even 
harder to solve computationally \citep{Vasquez2012,Nasser2013a}. By contrast, our models of 
population coupling are fully compatible with any model describing the dynamics of the population rate such as \cite{Mora2015}.

\bibliographystyle{natbib}

 % appendix_math_6
\appendix
\section*{Mathematical derivations}
\subsection{Derivation of the model form}
{\em Maximum entropy models.}
A maximum entropy model is defined by a distribution that maximizes its entropy,
\beq
S(P) = - \sum_\bs P(\bs) \; \text{log} P(\bs),
\eeq
while reproducing a set of chosen statistics. In the case where these statistics are the means of some observables $\mathcal{O}_1(\bs),\ldots,\mathcal{O}_M(\bs)$, the form of the model is given by:
\beq\label{maxent}
 P(\bs) = \frac{1}{Z} \exp\left( \sum_{a=1}^M \mu_a \, \mathcal{O}_a(\bs) \right),
\eeq
where $Z$ is a normalization factor. Eq.~\ref{maxent} is obtained by maximizing the entropy while constraining the chosen statistics using the method of Lagrange multipliers. 
The Lagrange multipliers $\mu_a$ are model parameters that must be adjusted so that the mean observables predicted by the model, $\langle \mathcal{O}_a \rangle_{\mu}$ agree with those of the data, $\langle \mathcal{O}_a \rangle_{\rm data} = (1/n)\sum_{\alpha=1}^n \mathcal{O}(\bs^{(\alpha)})$, where $(\bs^{(1)},\ldots,\bs^{(n)})$ are the $n$ activity patterns recorded in the experiment.
This fitting procedure is equivalent to maximizing the likelihood of the data under the model, $L=\prod_{\alpha=1}^n P(\bs^{(\alpha)})$, assuming that the patterns are independently drawn.
\CG{The likelihood maximization problem is convex and the distribution
  $P(\bs)$ maximizing the likelihood is always unique. However, if the
  constrained observables are linearly related, the optimal set of
  $\mu_a$ is not unique (even though the resulting distribution is), and must be set by chosing a convention.
}
\\

{\em Minimal model.}
In the minimal model, the statistics we constrain are $P(\s_i\!=\!1)$ for each neuron $i$, and $P(K\!=\!k)$ for each $k=0,...,N$. They correspond to the means of the following observables:
\begin{eqnarray}
P(\s_i\!=\!1) &=& \langle  \s_i   \rangle, \\
P(K = k) &=& \langle  \delta_{K,k}   \rangle,
\end{eqnarray}
where $\delta_{x,y}$ is Kronecker's delta, equal to 1 if $x\!=\!y$ , and 0 otherwise.
Note that while in the main text we use $K$ both as a short-hand for $\sum_i \s_i$ and its realization as a random variable, here we distinguish the two by using $K$ and $k$, respectively.
Applying Eq. \ref{maxent} to this choice of observables $(\sigma_i,\delta_{K,k})$ yields:
\begin{eqnarray}
P(\bs) &=& \dfrac{1}{Z} \exp\left( \sum_{k=0}^N \nu_k \delta_{K,k} + \sum_{i=1}^N \alpha_i \s_i \right) \\
&=& \dfrac{1}{Z} \exp\left( \nu_K + \sum_{i=1}^N \alpha_i \; \s_i  \right),
\label{eqn:intermediate_model}
\end{eqnarray}
where each $\nu_k$ is associated with the contraint on $\<\delta_{K,k}\>$ and each $\alpha_i$ with the constraint on $\<\s_i\>$.
In the second line we have used the fact that in the first sum, the only term which is non-zero is the one for which $k=K$. 

For convenience, we rescale the parameters $\nu_K$, which will give a common form to our three models. We first set $\nu_0 \!=\!0$, which is possible because the model is invariant when adding a constant to all $\nu_k$ (this only changes the normalization factor $Z$). We then intoduce the rescaled parameters $\beta_K$, defined as $\beta_0 \!=\!0$ and $\beta_K \!= \nu_K / \!K$ for $K>0$. We have $\nu_K = K\beta_K=\sum_i \s_i \beta_K $, so that the model takes the form : 
\beq
P(\bs) = \dfrac{1}{Z} \exp\left( \sum_{i=1}^N (\alpha_i+\beta_K) \; \s_i  \right).
\eeq
This model has $2N-1$ parameters: there are $N$ coefficients $(\alpha_i)_{i=1}^N$ and $N+1$ coefficients $(\beta_k)_{k=0}^N$, but $\beta_0$ is not used, and the model is invariant under a change in parameters $\alpha_i' = \alpha_i + c$, $\beta_k' = \beta_k -c$, for any number $c$. \\

{\em Linear-coupling model.}
The linear-coupling model reproduces $P(\s_i)$ and $P(K)$, and also the linear correlation between the neuron response $\s_i$ and the population rate $K$, $\<\sigma_i K\>$. The three sets of constrained observables are thus $(\s_i)_{i=1,\ldots,N}$, $(\delta_{K,k})_{k=0,\ldots,N}$, and  $(\s_i K)_{i=1,\ldots,N}$. With this choice of observables Eq. \ref{maxent} reads:
\begin{align}
P(\bs) &= \dfrac{1}{Z} \exp\left( \sum_{k=0}^N \nu_k \delta_{K,k} + \sum_{i=1}^N \alpha_i \s_i  + \sum_{i=1}^N \; \gamma_i K \s_i \right) \\
&= \dfrac{1}{Z} \exp\left( \nu_K + \sum_{i=1}^N ( \alpha_i + \gamma_i K) \; \s_i  \right),
\end{align}
where, in addition to the $\alpha_i$ and $\nu_k$ parameters, each $\gamma_i$ is associated with the constraint on $\<\s_i K\>$.
Note that in general the inferred parameters $\alpha_i$ and $\nu_k$ will be different from the ones inferred in the minimal model. This is due to the fact that the set of observables $\s_i$, $\delta_{K,k}$ and $\s_i K$ are not independent. Therefore the parameters $\gamma_i$ cannot be learned independently from $\alpha_i$ and $\nu_k$.

As for the minimal model, we rescale the parameters $\nu_K$ with $\beta_0=0$ and $\beta_K = \nu_K /K$ for $K>0$:
\beq
P(\bs) = \dfrac{1}{Z} \exp\left( \sum_{i=1}^N (\alpha_i+\beta_K + \gamma_i K) \; \s_i  \right).
\eeq

This model has $3N-2$ parameters: there are $2N$ coefficients $(\alpha_i)_{i=1}^N$ and $(\gamma_i)_{i=1}^N$ and $N+1$ coefficients $(\beta_k)_{k=0}^N$, but $\beta_0$ is not used, and the model is invariant under a changes in parameters $\alpha_i' = \alpha_i + c$, $\beta_k' = \beta_k -c +dK$, $\gamma_i'=\gamma_i - d$ for any numbers $c$ and $d$. \\

{\em Complete coupling model.}
The third maximum entropy model reproduces the joint probability distributions between the response of each neuron and the population rate, $P(\s_i, K)$. The problem reduces to matching $P(\s_i\!=\!1,K)$ for all $i=1,\ldots,N$ and $K=0,\ldots,N$, since $P(\s_i\!=\!0,K)$ can be determined through:
\beq
 P(\s_i\!=\!0,K) = P(K) -  P(\s_i\!=\!1,K),
\eeq
where the distribution $P(K)$ is set by:
\begin{equation}
 \sum_{i=1}^N P(\s_i\!=\!1,K) =  K P(K)
\end{equation}
This holds true because $K$ is the number of neurons spiking so:
\begin{equation}
 \sum_{i=1}^N P(\s_i\!=\!1|K)  = \sum_{i=1}^N \< \s_i | K\> =  \< \sum_{i=1}^N \s_i | K\>=  K
\end{equation}
where we can then mutliply both sides by $P(K)$. Here $\< \, . \, |K \>$ stands for the mean conditioned by $K$.

Therefore, we impose that the model only reproduces the statistics $P(\s_i\!=\!1,K)$, which are the means of the observables $\s_i\delta_{K,k}$. Using Eq.~\ref{maxent} with this set of observables yields:
\begin{align}
 P(\bs) &= \dfrac{1}{Z} \exp \left(\sum_{i=1}^N \sum_{k=0}^N h_{ik}\; \s_i \delta_{K,k}\right) \\
&= \dfrac{1}{Z} \exp \left(\sum_{i=1}^N h_{iK} \: \s_i \right),
\end{align}
where each $h_{ik}$ is associated with the constraint on $\<\s_i \delta_{K,k}\>$.

This model has $N(N\!-\!1)\!+\!1$ parameters: there are $N(N+1)$ coefficients $(h_{iK})_{i=1, K=0}^{N,N}$, but the $N$ coefficients $(h_{i0})_{i=1}^{N}$ are not used, and only the sum $\sum_i h_{iN}$ of the $N$ coefficients $(h_{iN})_{i=1}^{N}$ is used, when all neurons spike simultaneously.

\subsection{Regularization}
We regularized the empirical population rate distribution $P(K)$ and conditional firing rates $P(\s_i|K)$ using pseudocounts. If we denote by \CG{$n=2.8\, 10^5$} the total number of responses \CG{$\sigma$ recorded during} the experiment and by $n_K$ the number of responses with $K$ spikes in the population, the distribution of population rates $K$ was computed as:
\begin{equation}
 P(K)   =  \dfrac{n_\CG{K} + \lambda P_{\rm indep}(K)}{n + \lambda},
\end{equation}
where $P_{\text{indep}}(K)$ is the distribution of $K$ in a model of independent neurons reproducing the empirical firing rates $\<\sigma_i\>$. Similarly, if we denote by $n_{iK}$ the number of responses in which neuron $i$ spiked and in which the population rate was $K$, the conditional firing rates were estimated as:
\begin{equation}
 P(\s_i=1|K) = \dfrac{ n_{iK} + \lambda P_{\rm indep}(\s_i=1|K)}{n_{K} + \lambda},
\end{equation}
where again $P_{\rm indep}(\s_i=1|K)$ is the estimate of the conditional firing rate according to the independent model. The terms scaling as $\lambda$ play the role of pseudocounts. These pseudocounts are not taken to be uniform, but rather follow the prediction of a model of independent neurons.
We used $\lambda=1$ so that the total weight of pseudocounts is equivalent to a single observed pattern.

\subsection{Calculating statistics from the model}
We start by providing an analytical expression for the normalization factor, defined as:
\beq
Z = \sum_{\bs} \exp\left( \sum_{i=1}^N h_{iK} \s_i \right).
\eeq
All useful statistics predicted by the model can be derived from the expression of $Z$, as we shall see below.
To calculate $Z$, we decompose it as a sum over groups of patterns with the same population activity $K$: $Z=\sum_{k=0}^N Z_k$ with:
\begin{align} \label{eq:def_Qk}
Z_k &= \sum_{\substack{\bs\\ K = k}} \text{exp}\left(\sum_{i=1}^N h_{iK} \s_i\right) \\
&= \sum_{i_1 < ... < i_{k}} \text{exp}\left(\sum_{b=1}^k h_{i_b,k}\right)
\end{align}
We introduce the polynomial $Q(X) = \prod_{i=1}^N(1+e^{h_{ik}}X)$. Expanding $Q$, we can calculate its coefficient of order $X^k$, denoted by Coeff$[Q,X^k]$. This coefficient is the sum of all the terms having exactly $k$ factors $e^{h_{ik}}$:
\begin{align}
\text{Coeff}[Q,X^k] &= \sum_{i_1 < ... < i_{k}} \; \prod_{b=1}^k \exp\left({h_{i_bk}}\right) \\
&= \sum_{i_1 < ... < i_{k}} \exp\left(\sum_{b=1}^k {h_{i_bk}}\right) \\
&= Z_k
\end{align}
It is obtained by recursively computing the coefficients of $\prod_{i=1}^n(1+e^{h_{ik}}X)$ of order up to $X^k$,  for $n=1$ to $N$, using the relation:
\begin{equation}
 \text{Coeff}[(1+bX)F,X^l] = \text{Coeff}[ F,X^l] + b\, \text{Coeff}[ F,X^{l-1}],
\end{equation}
for any number $b$, polynomial $F$ and order $X^l$. $Z_k$ can thus be computed in time linear in $kN$, and $Z=\sum_k Z_k$ can be computed rapidly.

 Many statistics of the model can then be calculated by deriving $Z$. For example, the mean observables according to the model in Eq. \ref{maxent} are given by:
 \beq \label{maxent_mean}
\<\mathcal{O}_a\>_\mu = \dfrac{ \partial \log Z }{ \partial \mu_a}
 \eeq
This formula gives the following expression for the joint distribution of  $\sigma_i$ and $K$: 
\begin{align}
P(\sigma_i\!=\!1,K)=&\dfrac{ \partial \log Z }{ \partial h_{iK}}\\
=&\frac{1}{Z}\mathrm{Coeff}\!\!\left[Xe^{h_{iK}}\prod_{j\neq i}(1+Xe^{h_{j,K}}),X^K\right]. \label{PsigmaK}
\end{align}
 Similarly, pairwise correlations are computed using the formula:
\beq
\<\sigma_i\sigma_j\>=\frac{1}{Z}\sum_{K}\mathrm{Coeff}\!\!\left[X^2e^{h_{iK}+h_{jK}}\prod_{l\neq i,j}(1+Xe^{h_{lK}}),X^K\right].
\eeq

\subsection{Model inference}
To learn the model parameters from the data, we maximized the normalized log-likelihood $\mathcal{L} = (1/n) \log L$ using Newton's method. The update equation for the parameter values in Newton's method read:
\begin{equation}
{\bm \mu}^{(t+1)} = {\bm \mu}^{(t)} - a \, {\mathcal{\bm H}}^{-1} \cdot {\bm \nabla} \mathcal{L},
\end{equation}
where $a$ is an adjustable step size taken typically between 0.1 and 1. $\bm \mu^{(t)}=(\mu_a^{(t)})_{a=1,\ldots,M}$ is the vector of the parameters at iteration $t$;
${\bm \nabla} \mathcal{L}$  and $\mathcal{H}$ are the gradient and Hessian of $\mathcal{L}$ with respect to the parameters $\mu_a$. In the general context of maximum entropy models (Eq.~\ref{maxent}), one can show that the gradient and Hessian read:
\begin{eqnarray}
({\bm \nabla} \mathcal{L})_a=\frac{\partial\mathcal{L}}{\partial \mu_a} &=& \langle \mathcal{O}_a \rangle_{\rm data} - \langle \mathcal{O}_a \rangle_{\mu}\\ 
&=& \langle \mathcal{O}_a \rangle_{\rm data}-\frac{\partial\log Z}{\partial \mu_a}  \label{grad_2} \\
 \mathcal{H}_{ab}=\frac{\partial^2\mathcal{L}}{\partial \mu_a \partial \mu_b} &=& \langle \mathcal{O}_a \rangle_{\mu} \langle \mathcal{O}_b\rangle_{\mu} - \langle \mathcal{O}_a \mathcal{O}_b \rangle_{\mu}  \\ 
&=& -\frac{\partial^2\log Z}{\partial \mu_a \partial \mu_b}. \label{grad_4}
\end{eqnarray}
where we used Eq. \ref{maxent_mean} for Eq. \ref{grad_2} and a similar formula for Eq. \ref{grad_4}. Both quantities can readily be computed as derivatives of the normalization factor $Z$.

For time efficiency, we only updated the Hessian every 100 iterations
of the algorithm. 
We stopped the algorithm when the fitting error reached $10^{-6}$.
The fitting error was defined as the maximum error on $P(K)$ and $P(\s_i)$ for the minimal model, on $P(K)$, $P(\s_i)$ and $\langle K\s_i\rangle$ for the linear-coupling model, and on $P(K)$ and $P(\s_i|K)$ for the complete coupling model.

 The code for the models inference is available at \url{https://github.com/ChrisGll/MaxEnt_Model_Population_Coupling}.

\end{document}